\documentclass[aps,prd,12pt,preprint,nofootinbib,tightenlines,floatfix,superscriptaddress,showpacs]{revtex4}

\usepackage{graphicx}% Include figure files
\usepackage{rotating}
\usepackage{dcolumn}% Align table columns on decimal point
\usepackage{bm}% bold math
\usepackage{longtable}% Include figure files

\begin{document}

\def\figwid{6.5in}

\newcommand{\ccb}{{\psi(nS)}}
\newcommand{\qqb}{{q\bar{q}}}
\newcommand{\eprtoeta}{{{\cal B}[\ccb\to\geta]/{\cal B}[\ccb\to\gepr]}}
\newcommand{\gp}{{\gamma P}}
\newcommand{\chgp}{{\ccb\to\gp}}

\newcommand{\ie}{{\it i.e.}}
\newcommand{\etal}{{\it et al.}}
\newcommand{\etap}{{\eta\,'}}
\newcommand{\etav}{{\eta^{(\prime)}}}
\newcommand{\etac}{{\eta_c}}
\newcommand{\psip}{{\psi(2S)}}
\newcommand{\psit}{{\psi(3770)}}
\newcommand{\diel}{{e^+e^-}}
\newcommand{\dimu}{{\mu^+\mu^-}}
\newcommand{\dipi}{{\pi^+\pi^-}}
\newcommand{\dipiz}{{\pi^0\pi^0}}
\newcommand{\jpsi}{{J/\psi}}
\newcommand{\piz}{{\pi^0}}

\newcommand{\gga}{{\gamma\gamma}}
\newcommand{\zze}{  {\dipiz\eta }}
\newcommand{\zzegg}{  {\zze(\gga) }}
\newcommand{\ppe}{  {\dipi\eta }}
\newcommand{\ppegg}{{\dipi\eta (\gga)}}
\newcommand{\tpi}{{\dipi\piz}}
\newcommand{\tpz}{{3\pi^0}}
\newcommand{\ppg}{{\gamma\pi^+\pi^-}}

\newcommand{\gpiz}{{\gamma\piz}}
\newcommand{\geta}{{\gamma\eta}}
\newcommand{\gepr}{{\gamma\etap}}
\newcommand{\getav}{{\gamma\etav}}
\newcommand{\omg}{{\gamma\omega(\tpi)}}

\newcommand{\eprppegg}{{\etap\to\dipi\eta(\gga)}}
\newcommand{\eprppeppg}{{\etap\to\dipi\eta(\ppg)}}
\newcommand{\eprzzegg}{{\etap\to \dipiz\eta(\gga)}}

\newcommand{\getatpz}{{\geta(\tpz)}}
\newcommand{\getatpi}{{\geta (\tpi)}}
\newcommand{\geprppegg}{{\gepr[\dipi\eta(\gga)]}}
\newcommand{\geprppetpz}{{\gepr[\dipi\eta(\tpz)]}}
\newcommand{\geprppetpi}{{\gepr [\dipi\eta(\tpi)]}}

\newcommand{\jpgpiz}{{\jpsi\to\gpiz}}
\newcommand{\ppgpiz}{{\psip\to\gpiz}}
\newcommand{\ptgpiz}{{\psit\to\gpiz}}

\newcommand{\myspace}{{\ \ \ }}

\newcommand{\jpgeta}{{\jpsi\to\geta}}
\newcommand{\ppgeta}{{\psip\to\geta}}
\newcommand{\ptgetav}{{\psit\to\getav}}
\newcommand{\ttgeta}{{\myspace\to\geta}}
\newcommand{\ttgetagg}{{\ttgeta (\gga)}}
\newcommand{\ttgetatpz}{{\ttgeta (\tpz)}}
\newcommand{\ttgetatpi}{{\ttgeta (\tpi)}}
\newcommand{\ttgetappg}{{\ttgeta (\ppg)}}
\newcommand{\ttgetaeeg}{{\ttgeta (\eeg)}}

\newcommand{\jpgepr}{{\jpsi\to\gepr}}
\newcommand{\ppgepr}{{\psip\to\gepr}}
\newcommand{\ptgepr}{{\psit\to\gepr}}
\newcommand{\ttgepr}{{\myspace\to\gepr}}

\newcommand{\jpgeprppegg}{{\jpgepr [\dipi\eta(\gga)]}}
\newcommand{\ppgeprppegg}{{\ppgepr [\dipi\eta(\gga)]}}
\newcommand{\ptgeprppegg}{{\ptgepr [\dipi\eta(\gga)]}}
\newcommand{\ttgeprppegg}{{\ttgepr [\dipi\eta(\gga)]}}

\newcommand{\jpgeprppetpz}{{\jpgepr [\dipi\eta(\tpz)]}}
\newcommand{\ppgeprppetpz}{{\ppgepr [\dipi\eta(\tpz)]}}
\newcommand{\ptgeprppetpz}{{\ptgepr [\dipi\eta(\tpz)]}}
\newcommand{\ttgeprppetpz}{{\ttgepr [\dipi\eta(\tpz)]}}

\newcommand{\jpgeprppetpi}{{\jpgepr [\dipi\eta(\tpi)]}}
\newcommand{\ppgeprppetpi}{{\ppgepr [\dipi\eta(\tpi)]}}
\newcommand{\ptgeprppetpi}{{\ptgepr [\dipi\eta(\tpi)]}}
\newcommand{\ttgeprppetpi}{{\ttgepr [\dipi\eta(\tpi)]}}

\newcommand{\jpgeprppeppg}{{\jpgepr [\dipi\eta(\ppg)]}}
\newcommand{\ppgeprppeppg}{{\ppgepr [\dipi\eta(\ppg)]}}
\newcommand{\ptgeprppeppg}{{\ptgepr [\dipi\eta(\ppg)]}}
\newcommand{\ttgeprppeppg}{{\ttgepr [\dipi\eta(\ppg)]}}
\newcommand{\geprzzegg}{{\gepr[\dipiz\eta(\gga)]}}
\newcommand{\ttgeprzzegg}{{\ttgepr [\dipiz\eta(\gga)]}}
\newcommand{\geprppg}{{\gepr (\ppg)}}
\newcommand{\ttgeprppg}{{\ttgepr (\ppg)}}
\newcommand{\ttgepromg}{{\ttgepr [\omg]}}
\newcommand{\ttgeprgg}{{\ttgepr (\gga)}}

\newcommand{\bjpgpiz}{{{\cal B}(\jpgpiz)}}
\newcommand{\etagg}{{\eta\to\gga}}
\newcommand{\etatpz}{{\eta\to 3\piz}}
\newcommand{\etatpi}{{\eta\to\tpi}}
\newcommand{\etappg}{{\eta\to\ppg}}
\newcommand{\etavppg}{{\etav\to\ppg}}
\newcommand{\etaeeg}{{\eta\to\diel\gamma}}

\newcommand{\eprppe}{{\etap\to\dipi\eta}}
\newcommand{\eprzze}{{\etap\to\dipiz\eta}}
\newcommand{\eprppg}{{\etap\to\dipi\gamma}}
\newcommand{\epromg}{{\etap\to\omg}}
\newcommand{\eprgg}{{\etap\to\gga}}

\newcommand{\ometpi}{{\omega\to\tpi}}
\newcommand{\chivdof}{\chi^2_V/\mathrm{d.o.f.}}
\newcommand{\chimdof}{\chi^2_E/\mathrm{d.o.f.}}

\preprint{CLNS~09/2050}
\preprint{CLEO~09-03}  

\title{\Large \boldmath\bf
Charmonium decays to $\gamma\pi^0$, $\gamma\eta$, and $\gamma\eta^{\,\prime}$\\
}

\author{T.~K.~Pedlar}
\author{J.~Xavier}
\affiliation{Luther College, Decorah, Iowa 52101, USA}
\author{D.~Cronin-Hennessy}
\author{K.~Y.~Gao}
\author{J.~Hietala}
\author{T.~Klein}
\author{R.~Poling}
\author{P.~Zweber}
\affiliation{University of Minnesota, Minneapolis, Minnesota 55455, USA}
\author{S.~Dobbs}
\author{Z.~Metreveli}
\author{K.~K.~Seth}
\author{B.~J.~Y.~Tan}
\author{A.~Tomaradze}
\affiliation{Northwestern University, Evanston, Illinois 60208, USA}
% ...
\collaboration{CLEO Collaboration}
\noaffiliation

\date{April 8, 2009}

\begin{abstract}

Using data acquired with the CLEO-c detector at the 
CESR $\diel$ collider, we measure branching fractions 
for $\jpsi$, $\psip$, and $\psit$ decays to 
$\gpiz$, $\geta$, and $\gepr$. Defining
$R_n\equiv\eprtoeta$, we obtain $R_1=(21.1\pm0.9)\%$
and, unexpectedly, an order of magnitude smaller limit, 
$R_2<1.8\%$ at 90\%~C.L. We also use $\jpgepr$ events 
to determine branching fractions of improved precision 
for the five most copious $\etap$ decay modes.

\end{abstract}


\maketitle

  In the conventional view of quantum chromodynamics (QCD),
the mass eigenstates $\jpsi$, $\psip$, and $\psit$ represent 
the $1S$, $2S$, and $1D$ $c\bar{c}$ states, respectively
(with some evidence of modest $2S-1D$ mixing~\cite{rosner2s1d}).
Most hadronic decays of these states that are not
transitions to lower-lying charmonia or decays to open charm 
can be described at tree level by 
$c\bar{c}$ annihilation into either three gluons ($ggg$) or a 
photon plus two gluons ($\gamma gg$).
However, some final states can be reached 
by less common routes, and their study
can lend experimental constraints on the relevant QCD predictions.
In this article we describe measurements of
branching fractions for each of $\jpsi$, $\psip$,
and $\psit$ decays to three such final states:
$\gpiz$, $\geta$, and $\gepr$.
In the $b\bar{b}$ system, limits of
${\cal B}(\Upsilon(1S)\to\getav)< 1.0 (1.9)\times 10^{-6}$ 
have been set~\cite{vijay} and are smaller than expected.
Comparable studies in the charmonium sector 
are warranted.

Vector charmonium decays to a photon and a 
light pseudoscalar ($\chgp$) can be described by a 
variety of mechanisms at the parton level, which
we label $(i)-(vi)$. When 
$P=\eta$ or $\etap$, 
the primary source is generally thought to be
$(i)~\ccb\to\gamma gg$, 
which then fragments to exclusive final states in a 
flavor-blind manner. The $\eta$ and $\etap$ mesons are commonly 
understood as mixtures of the pure SU(3)-flavor octet 
[$(u\bar{u}+d\bar{d}-2s\bar{s})/\sqrt{6}$]
and singlet [$(u\bar{u}+d\bar{d}+s\bar{s})/\sqrt{3}$] 
states and a small gluonium 
component~\cite{rosner,gilman}. The flavor content of each $\etav$
mass eigenstate can be quantified with a mixing angle, the value of
which becomes manifest in ratios of branching fractions
for various radiative decays involving an $\eta$ or 
$\etap$~\cite{BESJP,kloemix,escrib}.
Under the assumption that all $\ccb\to\getav$ final states 
occur through $(i)$, the $\eta-\etap$ mixing angle can be extracted 
from a measurement of $R_n\equiv\eprtoeta$ and compared with 
values obtained from other decays~\cite{BESJP,BESPP1}. 

  Alternate $c\bar{c}$ annihilation mechanisms for $\gp$ final states
are mediated either by three gluons or a virtual photon: 
$(ii)~\ccb\to ggg\to \qqb\gamma_{\rm fsr}$
(where $\gamma_{\rm fsr}$ represents final state radiation
off one of the quarks) or
$(iii)~\ccb\to\gamma^*\to q\bar{q} \gamma_{\rm fsr}$.
It has also been suggested that an M1 transition to $\etac$ 
followed by $\etac$-$\etav$ mixing~\cite{frijack,ktchao}
$(iv)~\ccb\to\gamma\etac^*\to\getav$ could contribute.
The vector dominance model (VDM) can be used to predict~\cite{intem}
$(v)~\psip\to\jpsi^*\eta\to\geta$.
For all these processes $R_1\approx R_2$ is expected.
Previous measurements~\cite{PDG2008} 
yield $R_1 = (20.8\pm2.4)\%$, 
consistent with the dominance of $(i)$ with an $\eta$-$\etap$ mixing angle
in the expected range~\cite{PDG2008}. However, rate determinations 
for $\psip$ decays have yielded $R_2<66\%$ at 90\%\
confidence level (C.L.)~\cite{PDG2008} and hence are not 
yet precise enough to confirm a value comparable to $R_1$. 
No measurements for
$\ptgetav$ have yet been reported. 

When $P=\piz$, processes $(i)$ and $(iv)$ violate isospin conservation,
and are therefore suppressed relative to
process $(iii)$ or $(v)$ above, or $(vi)~ggg\to 2(q\bar{q})\to\rho^{0*}\piz$,
$\rho^{0*}\to\gamma$ via VDM.
Ref.~\cite{cherny} (CZ) finds process
$(ii)$ to be negligible, but 
$(iii)$ and $(vi)$ to be of 
comparable magnitude and fully coherent.
Updating the CZ VDM calculation to current
experimental branching fractions and widths~\cite{PDG2008} gives
$\bjpgpiz\approx 6\times 10^{-5}$, nearly double
the PDG average measurement of 
$(3.3^{+0.6}_{-0.4})\times 10^{-5}$~\cite{PDG2008}.
A similar disparity was measured in Ref.~\cite{cleogp} 
for $\gamma^* \gamma\to\piz$ for spacelike non-asymptotic 
momentum transfers in the range
$|q^2| = 1.5 - 9$ GeV$^2$, where the CZ prediction was found to 
significantly overshoot the data.
Further experimental precision on $\bjpgpiz$ would allow
a more precise 
indirect measurement of the $\gamma^*-\gamma\piz$ vertex of 
process~$(iii)$ for timelike photons~\cite{newros}.

 For this measurement, events were acquired 
at the CESR $e^+e^-$ collider with the CLEO detector~\cite{CLEO}, 
mostly in the CLEO-c configuration (5\% of $\psip$ data 
were collected with CLEO~III). The data samples comprise
$(27.4 \pm 0.6) \times 10^6$~\cite{xnext} 
produced $\psip$ mesons, of which
$(9.589 \pm 0.020 \pm 0.070) \times 10^6$~\cite{xnext}
decay into $\dipi\jpsi$
(our source for $\jpsi$ decays),
and 814~pb$^{-1}$ at $\sqrt{s}=3.773$~GeV,
corresponding to $(5.3\pm 0.1)\times 10^6$~\cite{dhad} 
produced $\psit$ mesons. 
We also utilize 20.7~pb$^{-1}$
of data taken at $\sqrt{s}=3.671$~GeV, just below
the $\psip$ peak, for a ``continuum'' subtraction
of backgrounds.

  We select events in the decay modes shown in Table~\ref{tab:tableyield};
modes that are not shown have sensitivity inferior to those we employ.
Every particle in each mode's decay chain must be found. 
Each such particle is constrained to originate 
from a single point (vertex) consistent with the measured beam spot.
We then constrain the sum of all four-momenta 
to the known $\psip$ mass~\cite{PDG2008} 
and initial $\diel$ three-momentum.
The vertex  and full event four-momentum 
kinematic fits must satisfy quality requirements of
$\chivdof<10$ and $\chimdof<10$, respectively,
which typically retain 90-99\% of signal events.
For $\ppgpiz$, tighter restrictions of
$\chivdof<3$ and $\chimdof<3$ suppress QED background.
Further selections are based on
four-momenta from the fit.

Masses of final state or intermediate mesons must be found in the
signal or sideband windows given in Table~\ref{tab:tablemass}.
Note that different windows apply depending on the
$\etav$ decay mode because mass resolution depends upon the final state.
The tracking system must find
zero, two, or four tracks of net charge zero for the $\piz/\eta/\etap$
decay products, and, for $\jpsi$ modes, exactly two additional
oppositely-charged particles for the $\psip$-to-$\jpsi$-transition.
Photon candidate showers must lie in the barrel region of the
calorimeter ($|\cos\theta|<0.75$) where the
energy resolution and finding-efficiency are most favorable
and well-modeled.
Photon candidates must also have energy $E_\gamma >37$~MeV, lie more
than 30~cm from the center of any shower associated 
with one of the charged pions, and not align with the initial 
three-momentum of any $\pi^\pm$ candidate within 100~mrad.
A photon candidate is included in the decay chain only
if all more energetic ones are also used.
Photon pairs from $\piz$ or $\eta$  
are constrained to the appropriate mass~\cite{PDG2008},
except in $\etatpz$ and $\eprzzegg$; 
these final states are only required to have six photons with
a combined invariant mass in the corresponding mass window 
in Table~\ref{tab:tablemass}.
Photons in $\etavppg$ and $\eprppeppg$
must have $E_\gamma>100$~MeV to suppress feedacross from
other $\eta$ and $\etap$ decays; photons
in $\ppgpiz$ and  $\ptgpiz$ 
must have $E_\gamma>300$~MeV to suppress $\diel\to 3\gamma$.

For $\psit$ modes, where signal yields are expected to be small,
backgrounds larger, and systematic considerations much less important,
we take measures to enhance efficiency and suppress backgrounds
at the expense of systematic precision. The photon-finding criteria 
are loosened so as to increase efficiency: barrel and endcap showers are
used ($|\cos\theta|$$<$0.83 or 0.85$<$$|\cos\theta|$$<$0.93), and showers
between 15 and 30~cm from a shower associated with a charged track
are accepted if they have a photon-like lateral shape.
In addition, the four-momentum constrained fit quality
is tightened to $\chimdof<3$ to suppress backgrounds.

  Efficiencies for signal and feedacross from other modes considered 
here are modeled with Monte Carlo (MC) samples that were generated using the
{\sc EvtGen} event generator~\cite{evtgen},
fed through a {\sc Geant}-based~\cite{geant} detector simulation,
and subjected to the same event selection criteria as the data. 
We model $\etagg$ and $\tpz$ as phase-space-like decays. 
The decay $\eta\to\ppg$ is simulated as mediated by a $\rho^0\to\dipi$, 
weighted with a factor $E_\gamma^3$, where $E_\gamma$
is the photon energy in the $\eta$ center-of-mass system,
which results in the excellent agreement with the measured
dipion mass distribution seen in Ref.~\cite{etabr}. 
The decay $\etap\to\ppg$ is handled similarly,
but using a factor of $E_\gamma^2$, and
MC-data agreement is shown in Ref.~\cite{metap}. 
We generate $\etatpi$ according to the
distribution measured in Ref.~\cite{layter}.
All other $\etap$ decays are generated using phase space.
For each decay mode, we try all combinations for assignment of the
selected tracks and showers (although in practice more than a single
successful one per event is rare).

  We also tally an event satisfying the selections if instead
of lying inside the light meson signal mass window it has
mass within one of the two sideband regions indicated in 
Table~\ref{tab:tablemass}. Similarly, the number of events
with $\dipi$ recoil mass inside the sidebands is accumulated
for each $\jpsi$ mode. To allow for possible double-counting
of non-$\jpsi$ and light meson mass-sideband backgrounds,
we only subtract half of the window-size-scaled dipion
recoil mass sideband number for each mode, and 
assign a systematic uncertainty of the full scaled number.

We subtract feedacross attributable to any radiative 
$\jpsi$ or $\psip$ decays into $\gpiz$, $\geta$, and $\gepr$, 
including the decay modes not selected. To do so, we employ
MC samples that are normalized to the observed net yields
and $\etap$ relative branching fractions found here,
except for $\eta$, $\omega$, and $\piz$ decay modes not probed, 
for which
PDG~\cite{PDG2008} branching fractions are used.
Continuum data counts satisfying the selections are
subtracted after scaling by relative
luminosity and $1/s$. The mass sideband subtraction is performed
in the MC and continuum samples, assuring that
the efficiency, feedacross, and continuum backgrounds
are computed correctly. This procedure implicitly
assumes that backgrounds other than from continuum
and feedacross are linear in pseudoscalar candidate mass;
it explicitly accounts for nonlinear backgrounds from
continuum and feedacross.

  Applying the above-described selection criteria to 
our data and MC samples yields the results shown in
Table~\ref{tab:tableyield}, in which raw yields
for signal and sidebands are given along with
appropriately scaled feedacross and continuum corrections.
The signal efficiency is also given for each mode,
which ranges from 5\% to 36\%. All modes except $\ppgpiz$
and $\ptgpiz$ have very small backgrounds.
No statistically significant signals are seen for $\ppgeta$,
$\ppgpiz$, or $\ptgpiz$.

\begin{figure}[tbh]
\begin{center}
\includegraphics[width=\figwid]{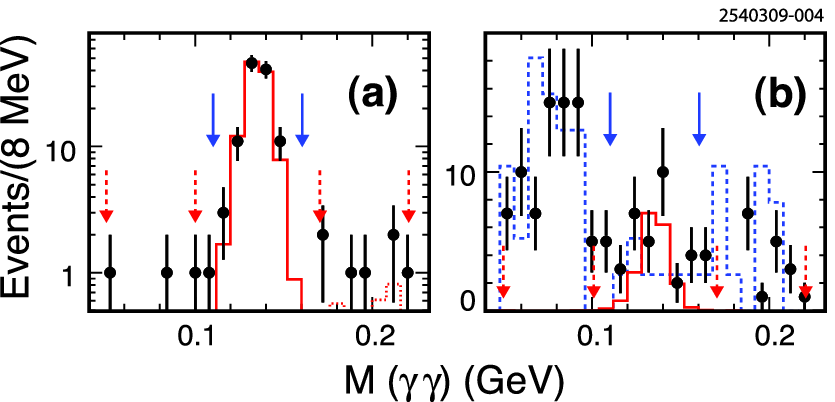}
\caption{Two-photon mass distributions for
$\gpiz$ final states from (a) $\jpsi$ and (b) $\psip$. 
Points show the $\psip$ data; 
dashed histogram, the luminosity-scaled continuum data;
solid line histogram, signal MC;
and dotted histogram, MC feedacross from other $\gp$ decays.
Upper solid arrows show nominal selection criteria; all selection
criteria are in place except those upon plotted mass.
Lower dashed arrows show sidebands.} 
\label{fig:pi0}
\end{center}
\end{figure}

\begin{figure}[tbh]
\begin{center}
\includegraphics[width=\figwid]{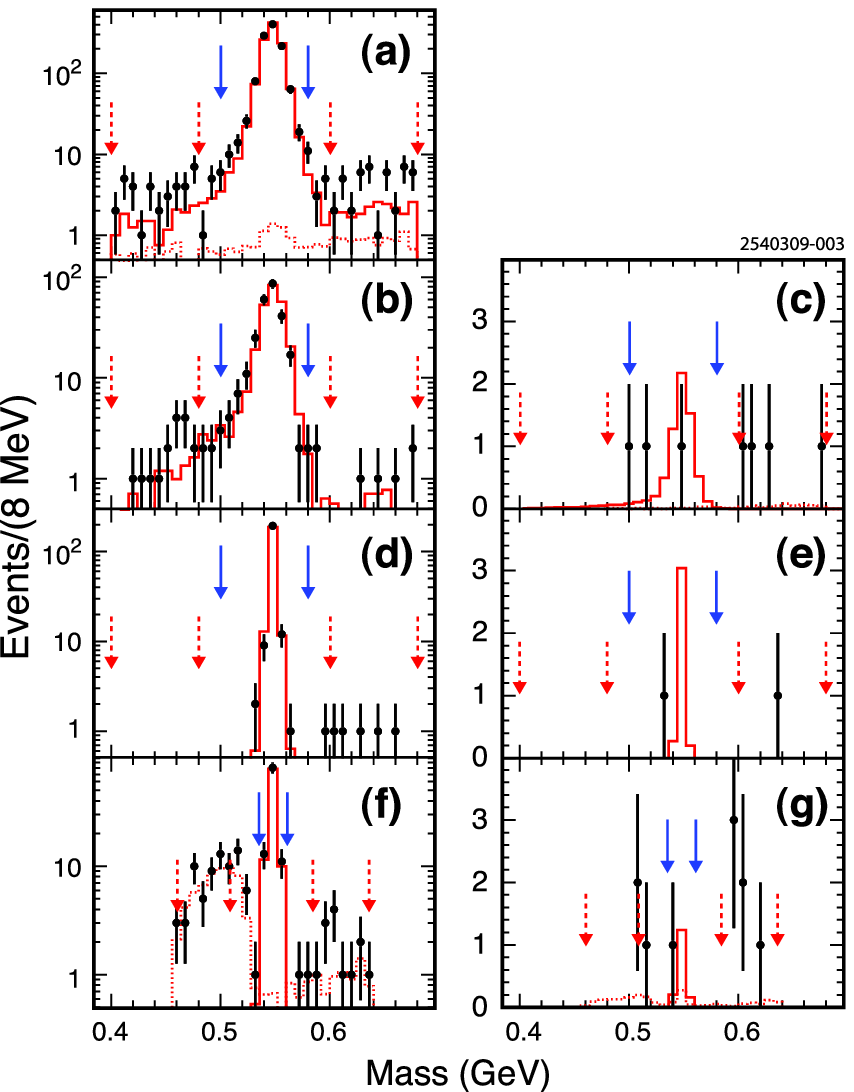}
\caption{Distributions of $\eta$ candidate mass for
$\geta$ final states from $\jpsi$ (left) and $\psip$ (right).
The $\eta$ decay modes are 
(a) $\gga$; 
(b), (c) $\tpz$;
(d), (e) $\tpi$;
and (f), (g), $\ppg$.
Symbols are defined in Fig.~1.}
\label{fig:eta}
\end{center}
\end{figure}

\begin{figure}[tbh]
\begin{center}
\includegraphics[width=5.0in]{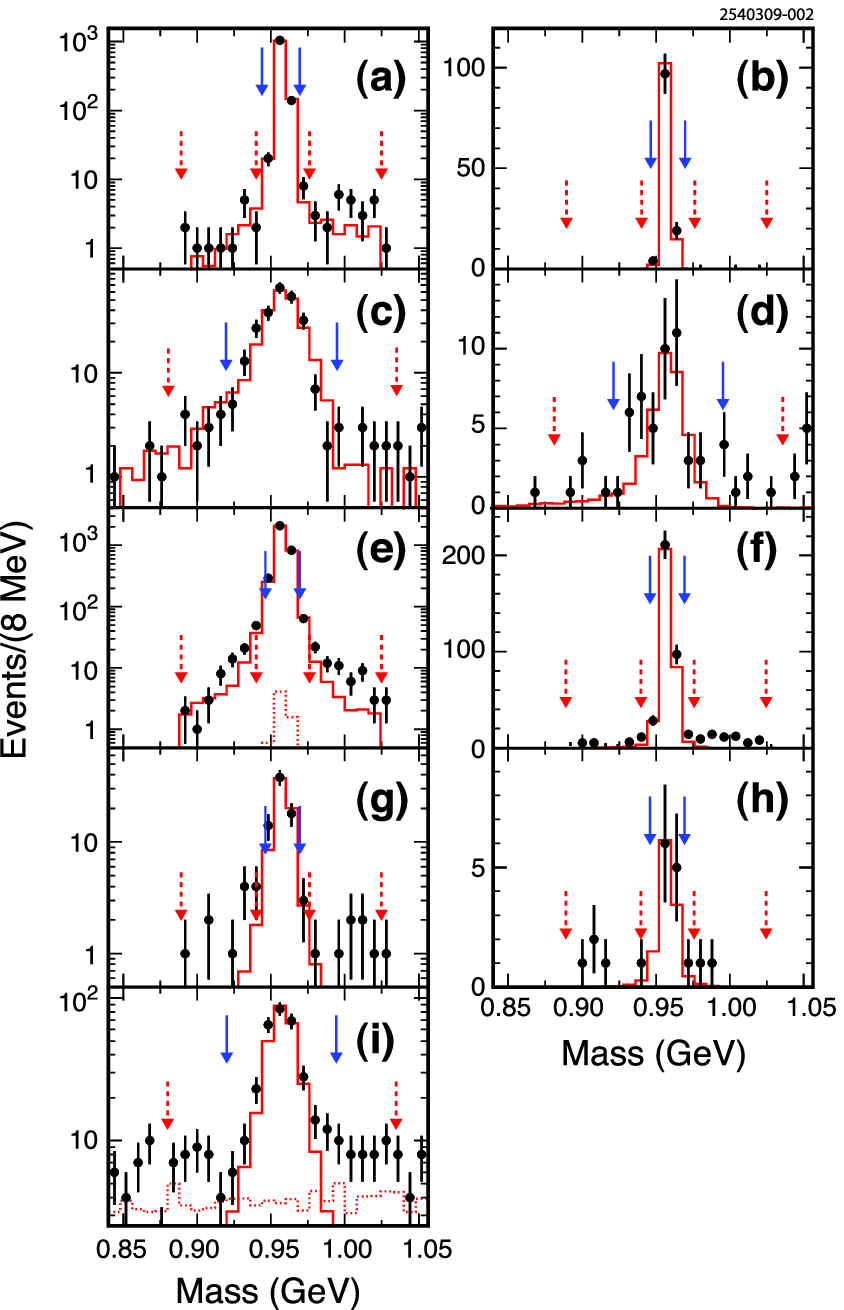}
\caption{Distributions of $\etap$ candidate mass for
$\gepr$ final states from $\jpsi$ (left) and $\psip$ (right).
The $\etap$ decay modes are (a), (b) $\ppegg$; 
(c), (d) $\zzegg$;
(e), (f) $\ppg$;
(g), (h) $\omg$;
and (i) $\gga$.
Symbols are defined in Fig.~1.
In (c), (d), and (i), only parts
of the sideband intervals are shown.}
\label{fig:epr}
\end{center}
\end{figure}

\begin{figure}[tbh]
\begin{center}
\includegraphics[width=\figwid]{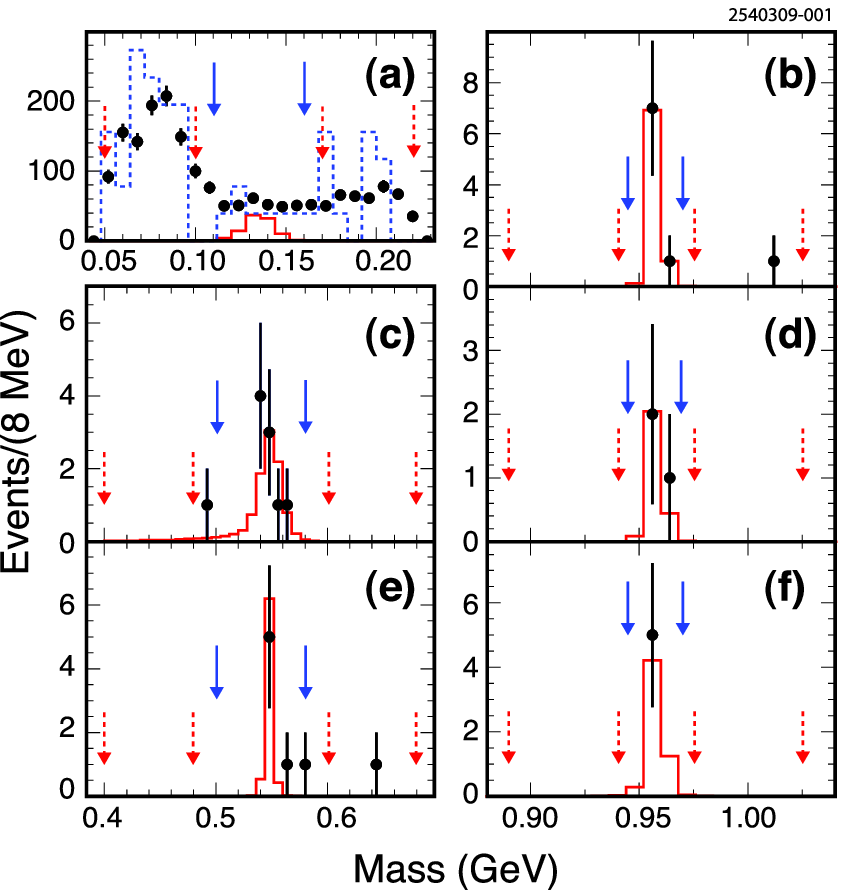}
\caption{Distributions of light quark meson candidate mass for
$\sqrt{s}=3.773$~GeV 
final states of 
(a) $\gpiz$;
(c), (e) $\geta$; and
(b), (d), (f) $\gepr$.
Decay modes are
(b) $\geprppegg$,
(c) $\getatpz$,
(d) $\geprppetpz$,
(e) $\getatpi$, and
(f) $\geprppetpi$.
Symbols are defined in Fig.~1.}
\label{fig:pdp}
\end{center}
\end{figure}

  The light meson mass distributions for all modes
are shown in Figs.~\ref{fig:pi0}-\ref{fig:pdp}; plots
for $\jpsi$ (other) modes have logarithmic (linear) vertical scales.
Background from $e^+e^-\to 3\gamma$ that contaminates
$\ppgpiz$ and $\ptgpiz$, visible in 
Figs.~\ref{fig:pi0}(b) and \ref{fig:pdp}(a), respectively, 
is irregular in shape but modeled well by the scaled
continuum data. There is also
a nonlinear background shape for modes involving
$\eta\to\gamma \dipi$ decays, but it is well-modeled
by the MC as it is due to feedacross from $\eta\to\tpi$.
A typical $\psip$-to-$\jpsi$-transition 
dipion recoil mass distribution appears
in Ref.~\cite{metap}, showing a tiny and flat
non-$\jpsi$ background; this is typical of all modes.
Ample signals for $\jpgpiz$, $\jpgeta$, $\jpgepr$,
and $\ppgepr$ are apparent and well-modeled by the MC-predicted 
shapes, aside from very small backgrounds that are
approximately linear in mass.

We make the first observations of $e^+e^-\to\getav$ at $\sqrt{s}=3.773$~GeV
with statistical significances of $5.0(6.4)\sigma$ and
observed cross sections $(0.17^{+0.05}_{-0.04}\pm0.03)$~pb ($\geta$) and
$(0.21^{+0.07}_{-0.05}\pm0.03)$~pb ($\gepr$).
The fraction of these values attributable to continuum cannot
be estimated accurately from our $\sqrt{s}=3.671$~GeV dataset 
due to its relatively small size: zero observed $\getav$ 
events yields upper limits at 90\%~C.L.~of $<2$~pb for both 
proceses. However, continuum cross sections for 
$e^+e^-\to\getav$ at $\sqrt{s}=10.58$~GeV have been measured 
by {\sc BaBar}~\cite{babargp} to be 
$4.5^{+1.2}_{-1.1}\pm0.3~(5.4\pm0.8\pm0.3)$~fb. 
Extrapolating~\cite{newros} to the charmonium
energy region requires knowledge of the $q^2$ dependence
of the form factors~\cite{brod} $F_{\etav}(q^2)$
for $\gamma^*\to\getav$. According to CLEO measurements~\cite{cleogp}
of the $\gamma\gamma^*\to\etav$ spacelike form factors, 
$|q^2 F_\etav(q^2)|$ is $15\pm 10~(5\pm 5)\%$ smaller than 
that at $q^2=m^2[\Upsilon(4S)]$
for $q^2$ in the charmonium region.
Continuum cross sections for $e^+e^-\to\getav$
are expected~\cite{brod} to scale as $|F_\etav(q^2)|^2$, leading
to an estimate of $0.19\pm0.07~(0.25\pm0.05)$~pb at $\sqrt{s}=3.773$~GeV.
These levels indicate a dominant continuum component in our signals.
Other potential sources of background for the
$\psit$ data sample are decays from the tail of
the $\psip$, $e^+e^-\to\psip\to\gp$ at
$\sqrt{s}=3.773$~GeV, and radiative returns (rr)
to the peak of the $\psip$,  
$e^+e^-\to\gamma_{\rm rr}\psip\to\gamma_{\rm rr}\gp$.
Using the methods of Ref.~\cite{tail},
these contributions are estimated
to be negligible: the former due to its inherently small rate
and the latter because such events will fail
the kinematic selections due to the presence of
$\gamma_{\rm rr}$.

Notable in Fig.~\ref{fig:eta}(c), \ref{fig:eta}(e),
and \ref{fig:eta}(g) is the absence of any $\ppgeta$ 
signal; we set an upper limit on the cross section at 
$\sqrt{s}=3.686$~GeV of $<1.1$~pb, about one tenth of
that expected if $R_2=R_1$ were true. 
We can rule out destructive interference with a continuum
$\geta$ signal as the primary cause of the deficit
because, as discussed above, the continuum
cross section is too small.

 The statistical errors on efficiencies
and subtractions for sideband, feedacross, and continuum
are combined with the statistical uncertainties on
event counts in the data. For systematic uncertainties associated
with finding tracks and showers, we assign
0.3\% per track and 0.4\% per photon~\cite{xnext}
on a mode-by-mode basis, accounting for
correlations in such uncertainties in
arithmetic computations involving two or more 
modes. For the looser photon criteria applied to $\psit$ modes,
a larger uncertainty of 3\% per photon is assigned.
Systematic uncertainties of
1\% (relative) are assigned to efficiencies,
uncorrelated mode-to-mode, and account
for any dataset or trigger modeling
dependence. We use $\eta$ branching fractions
and uncertainties from the PDG fit~\cite{PDG2008}.
As a systematic cross-check, we
compare ratios of $\jpgeta$ corrected yields
in different decay modes and find consistency
with our previous measurement~\cite{etabr} and 
Ref.~\cite{PDG2008}.

The net event yield for each channel listed
in Table~\ref{tab:tableyield} is divided by 
its respective efficiency, intermediate 
decay branching fraction (from PDG~\cite{PDG2008} 
for $\eta$ and this analysis for $\etap$), 
and total number of charmonia produced.
The resulting branching fractions appear
in Table~\ref{tab:tablefinal}, where 
multiple $\etav$ sub-mode values have been
combined in a weighted average.
For $\jpgpiz$, $\geta$, $\gepr$, and $\ppgepr$
the results are consistent with previous measurements.
We set improved upper limits for $\ppgpiz$ and $\geta$ 
and first upper limits for $\ptgpiz$, $\geta$, and $\gepr$.
These upper limits are computed using simulated trials
in which Poisson pseudo-random numbers are thrown for background and signal
levels, accounting for sideband-window-size scaling,
systematic errors, and fluctuations in the observed backgrounds, 
in a manner similar to that in Ref.~\cite{feldcous}.

The selected $\jpgepr$ events allow
measurement of 
branching fractions for the five
most common $\etap$ decay modes.
Ratios of branching fractions follow
from the $\jpgepr$ entries in 
Table~\ref{tab:tableyield} and the world average
value for ${\cal B}(\eta\to\gga)$~\cite{PDG2008}.
The resulting ratios
appear in Table~\ref{tab:tablebr} and represent
the most precise individual measurements~\cite{PDG2008}.
Absolute branching fractions for these five modes
can also be obtained by constraining their sum.
After combining  
branching fractions for $\tpi$ and $\dipi\diel$~\cite{rare}
with those for $\tpz$ and $\dimu\gamma$~\cite{PDG2008}, 
 $(99.2\pm0.2)$\% of all $\etap$ decays remain for
the largest five decay modes. 
We impose this constraint, accounting for
correlated systematic uncertainties, and
obtain the absolute branching fractions in Table~\ref{tab:tablebr}. 
The values shown are
consistent with those from the PDG 
constrained fit~\cite{PDG2008}
and of comparable precision.

In summary, we have described improved measurements of
and new limits on
branching fractions for charmonium decays to
a photon and a light pseudoscalar meson.
We also have performed the first simultaneous
measurement of the five largest $\etap$ 
branching fractions, attaining improved precisions.
Our result for $\bjpgpiz$ improves precision and
confirms earlier central
values~\cite{PDG2008}, yielding a better experimental
constraint~\cite{newros} on the $\gamma^*\to\gamma\piz$ vertex in the
timelike regime at $|q^2|=m^2(\jpsi)$.
For the ratio of $\eta : \etap$ production rates from each
resonance we obtain $R_1=(21.1\pm0.9)\%$ 
for $\jpsi$ and
$R_2<1.8\%$ at 90\%~C.L.~for $\psip$, where for both
results statistical and systematic uncertainties from the input branching
fractions from Table III have been combined in quadrature
after accounting for correlations.
Such a small value of $R_2$ is
unanticipated, posing a challenge to our understanding
of $c\bar{c}$ bound state decays.

We gratefully acknowledge the effort of the CESR staff
in providing us with excellent luminosity and running conditions.
D.~Cronin-Hennessy and A.~Ryd thank the A.P.~Sloan Foundation.
This work was supported by the National Science Foundation,
the U.S. Department of Energy,
the Natural Sciences and Engineering Research Council of Canada, and
the U.K. Science and Technology Facilities Council.

\clearpage

\begin{table}[th]
\setlength{\tabcolsep}{0.30pc}
\catcode`?=\active \def?{\kern\digitwidth}
\caption{For each decay mode, the number of events in the signal region ($N$);
the total number in sideband intervals, both for meson candidate mass ($S_M$) 
and, where applicable, transition 
dipion recoil mass ($S_J$), without scale factors
applied; the sum of scaled continuum and MC feedacross backgrounds ($F$),
which is negative when scaled sidebands exceed signal counts;
and the MC efficiency ($\epsilon$).}
\label{tab:tableyield}
\begin{center}
\begin{tabular}{lrrrrr}
\hline
\hline
Mode & $N$ & $S_M$ & $S_J$ & $F$ & $\epsilon$(\%) \\
\hline
\rule[10pt]{-1mm}{0mm}
$\jpgpiz$ & $    113$ & $      9$ & $     33$ & $     -2$ & $   31.1$ \\ 
$\ttgetagg$ & $   1137$ & $     80$ & $     69$ & $      4$ & $   26.3$ \\ 
$\ttgetatpz$ & $    256$ & $     21$ & $      6$ & $      0$ & $    7.0$ \\ 
$\ttgetatpi$ & $    217$ & $      5$ & $     30$ & $      0$ & $    9.0$ \\ 
$\ttgetappg$ & $    105$ & $     62$ & $     12$ & $    -12$ & $   21.0$ \\ 
$\ttgeprppegg$ & $   1208$ & $     36$ & $     56$ & $      3$ & $14.2$  \\ 
$\ttgeprzzegg$ & $    245$ & $     63$ & $     16$ & $     -2$ & $    4.9$  \\
$\ttgeprppg$ & $   3205$ & $    133$ & $    129$ & $      5$ & $   21.8$ \\ 
$\ttgepromg$ & $     71$ & $     18$ & $      8$ & $      0$ & $    5.6$ \\ 
$\ttgeprgg$ & $    317$ & $    230$ & $     16$ & $      0$ & $   22.7$ \\ 
$\ppgpiz$ & $     31$ & $     89$ &        -  & $    -33$ & $   20.0$   \\ 
$\ttgetatpz$ & $      2$ & $      4$ &        -  & $      0$ & $   15.8$ \\ 
$\ttgetatpi$ & $      1$ & $      1$ &        -  & $      0$ & $   12.6$ \\ 
$\ttgetappg$ & $      1$ & $      7$ &        -  & $     -1$ & $   32.1$ \\ 
$\ttgeprppegg$ & $    120$ & $      3$ &        -  & $      0$ & $   22.6$  \\ 
$\ttgeprzzegg$ & $     46$ & $     39$ &        -  & $      0$ & $   11.2$ \\ 
$\ttgeprppg$ & $    343$ & $     91$ &        -  & $      1$ & $   35.3$  \\ 
$\ttgepromg$ & $     12$ & $      6$ &        -  & $     -1$ & $    9.9$  \\ 

$\ptgpiz$ & $    331$ & $   1396$ &        -  & $   -468$ & $   19.8$ \\ 
$\ttgetatpz$ & $      9$ & $      0$ &        -  & $      0$ & $   19.4$ \\ 
$\ttgetatpi$ & $      7$ & $      1$ &        -  & $      0$ & $   23.2$ \\ 
$\ttgeprppegg$ & $      8$ & $      1$ &        -  & $      0$ & $   33.9$ \\ 
$\ttgeprppetpz$ & $      3$ & $      0$ &        -  & $      0$ & $    7.3$ \\ 
$\ttgeprppetpi$ & $      5$ & $      0$ &        -  & $      0$ & $   13.5$ \\ 
\hline
\hline
\end{tabular}
\end{center}
\end{table} 

\begin{table}[th]
\setlength{\tabcolsep}{0.55pc}
\catcode`?=\active \def?{\kern\digitwidth}
\caption{Mass windows and sideband (SB) intervals in MeV.
}
\label{tab:tablemass}
\begin{center}
\begin{tabular}{lccc}
\hline
\hline
Mode & Window & Lower SB & Upper SB \\
\hline
\rule[10pt]{-1mm}{0mm}
$\dipi$-recoil & 3087-3107 & 2980-3080 & 3114-3214 \\
$\piz\to\gga$ & 110-160 & 50-100 & 170-220 \\
$\eta\to\ppg$ & 535-560 & 460-510 & 585-635 \\
$\ \ \ \to$~other modes   & 500-580 & 400-480 & 600-680 \\
$\etap\to\gga,\ \zze$ & 920-995 & 730-880 & 1035-1185 \\
$\ \ \ \to$~other modes & 945-970 & 890-940 & 975-1025 \\
$\omega\to\dipi\piz$ & 750-814 & - & - \\
\hline
\hline
\end{tabular}
\end{center}
\end{table} 

\begin{table}[t]
\setlength{\tabcolsep}{0.28pc}
\catcode`?=\active \def?{\kern\digitwidth}
\caption{Branching fractions from this analysis, 
showing statistical and
systematic uncertainties, respectively,
and PDG~\cite{PDG2008}.
The rightmost column shows the difference
between the two in units of standard error ($\sigma$).
Upper limits are quoted at 90\%~C.L. 
Entries in the last two rows include the effects
of estimated continuum background and ignore
(include) maximal destructive interference
between $\psit$ and continuum sources.
}
\label{tab:tablefinal}
\begin{center}
\begin{tabular}{lrrr}
\hline
\hline
\rule[10pt]{-1mm}{0mm}
Mode & This result ($10^{-4}$) \ \  & PDG~($10^{-4}$) & \#$\sigma$ \\
\hline
\rule[10pt]{-1mm}{0mm}
$\jpgpiz$ & $     0.363 \pm      0.036 \pm      0.013$ & $      0.33^{+0.06}_{-0.04}$ & $  0.4$ \\ 
$\ttgeta$ & $     11.01 \pm       0.29 \pm       0.22$ & $       9.8 \pm        1.0$ & $  1.2$ \\ 
$\ttgepr$ & $      52.4 \pm        1.2 \pm        1.1$ & $      47.1 \pm        2.7$ & $  1.7$ \\ 
$\ppgpiz$ & $ <0.05$ & $       <54$ & -  \\ 
$\ttgeta$ & $ <0.02$ & $      <0.9$ & -  \\ 
$\ttgepr$ & $      1.19 \pm       0.08 \pm       0.03$ & $      1.36 \pm       0.24$ & $ -0.7$ \\ 
$\ptgpiz$ & $ <2$ & - & -  \\ 
$\ttgeta$ & $ <0.2~(1.5)$ & - & -  \\ 
$\ttgepr$ & $ <0.2~(1.8)$ & - & -  \\ 
\hline
\hline
\end{tabular}
\end{center}
\end{table} 

\begin{table}[th]
\setlength{\tabcolsep}{0.56pc}
\catcode`?=\active \def?{\kern\digitwidth}
\caption{Relative and absolute 
branching fractions for $\etap$
decays from this analysis, showing statistical and
systematic uncertainties.}
\label{tab:tablebr}
\begin{center}
\begin{tabular}{lrr}
\hline
\hline
\rule[10pt]{-1mm}{0mm}
Mode & ${\cal B}/{\cal B}(\ppe)$ (\%) & ${\cal B}$(\%)\ \ \ \ \ \ \\
\hline
\rule[10pt]{-1mm}{0mm}
$\dipi\eta$    & $\equiv 100$            & $ 42.4 \pm  1.1 \pm  0.4$ \\ 
$\dipiz\eta$   & $ 55.5 \pm 4.3 \pm 1.3$ & $ 23.5 \pm  1.3 \pm  0.4$ \\ 
$\ppg$         & $ 67.7 \pm 2.4 \pm 1.1$ & $ 28.7 \pm  0.7 \pm  0.4$  \\ 
$\gamma\omega$ & $ 5.5 \pm 0.7 \pm 0.1$  & $ 2.34 \pm 0.30 \pm 0.04$ \\ 
$\gga$         & $ 5.3 \pm 0.4 \pm 0.1$  & $ 2.25 \pm 0.16 \pm 0.03$ \\ 
\hline
\hline
\end{tabular}
\end{center}
\end{table}

\end{document}